\documentclass[%
 aip,
 amsmath,amssymb,
 reprint,%
]{revtex4-1}

\usepackage{graphicx}% Include figure files
\usepackage{dcolumn}% Align table columns on decimal point
\usepackage{xcolor}
\usepackage{bm}% bold math
\usepackage[utf8]{inputenc}
\usepackage[T1]{fontenc}
\usepackage{mathptmx}
\usepackage{etoolbox}
\usepackage[english]{babel}

\makeatletter
\def\@email#1#2{%
 \endgroup
 \patchcmd{\titleblock@produce}
  {\frontmatter@RRAPformat}
  {\frontmatter@RRAPformat{\produce@RRAP{*#1\href{mailto:#2}{#2}}}\frontmatter@RRAPformat}
  {}{}
}%
\makeatother
\begin{document}

\preprint{AIP/123-QED}

\title[Colliding rarefaction waves]{Collision of two radial rarefaction waves in unmagnetized ambient plasma:\\effects of the ambient plasma density}
% Force line breaks with \\
\author{M. François}
 \email{margaux.francois@u-Bordeaux.fr}
\affiliation{University of Bordeaux, CELIA, CNRS, CEA, F-33405 Talence, France}%

%\altaffiliation[Also at ]{Physics Department, XYZ University.}%Lines break automatically or can be forced with \\
\author{M. E. Dieckmann}%
\affiliation{ 
Link\"oping University, Department of Science and Technology (ITN), S-60174 Norrk\"oping, Sweden
}%

\author{L. Romagnani}
\affiliation{%
Sorbonne University, \'Ecole Polytechnique Paris, LULI, CNRS, CEA, F-91128 Palaiseau, France
}%

\author{X. Ribeyre}
\affiliation{%
CEA CESTA, F-33116 Le Barp, France
}%

\author{E. d'Humi\`eres}
\affiliation{University of Bordeaux, CELIA, CNRS, CEA, F-33405 Talence, France}%

\date{\today}% It is always \today, today,
             %  but any date may be explicitly specified

\begin{abstract}
The expansion of two circular rarefaction waves in vacuum or in a thin ambient plasma is examined with particle-in-cell simulations that resolve two spatial dimensions. In the simulation with no ambient plasma, the rarefaction waves interpenetrate near the symmetry line between both rarefaction wave centers. The exponential density decrease of rarefaction waves with distance implies that the sum of their density does not lead to a density maximum near the symmetry line. The absence of a density maximum, which would yield a repelling electric potential for the inflowing rarefaction wave ions near the symmetry line, and the high interpenetration speed of the ion beams lead to ion-ion instabilities rather than shocks in the overlap layer. The simulations with ambient plasma show that the rarefaction waves pile up the ions of the ambient plasma near the symmetry line. A localized piston of hot ambient ions forms. If its density is large enough, its thermoelectric field allows reverse shocks to grow in the rarefaction waves. These reverse shocks move slowly in the simulation frame and enclose a slab of downstream plasma. A decrease in the speed of the rarefaction wave ions upstream of the shocks with time leads to their collapse.
\end{abstract}

\maketitle

\section{\label{introduction}Introduction}

Ablating a solid target with an intense laser pulse creates a dense and energetic plasma near its surface. The small cross section of the laser beam ensures that this dense plasma is confined within a small volume. Driven by its thermal pressure, the outer layers of this dense plasma expand, forming a hemispherical rarefaction wave. The rarefaction wave front propagates into the dense plasma and accelerates its ions in the opposite direction. If the rarefaction wave formed by the accelerated ions expands freely, its density decreases exponentially with distance from the rarefaction wave front. In the presence of a density gradient, the thermal diffusion of electrons creates a charge imbalance. This charge imbalance induces what we call a thermoelectric field (sometimes called ambipolar electric field). It is the electric field that couples the fast expansion of electrons to the slow expansion of ions giving rise to an intermediate expansion speed. The thermoelectric field continues to accelerate the ions; the mean velocity of the accelerated ions increases linearly with distance from the rarefaction wave front\cite{Grismayer,Radial2}.

Secondary X-ray radiation from the ablated target ionizes the residual gas, which was trapped in the experimental chamber. The decrease in the rarefaction wave density with distance from the ablated target eventually leads to similar densities of the rarefaction wave plasma and the ambient plasma. At this distance, the ambient plasma starts to slow the rarefaction wave's expansion. The ambient plasma's density, and thus its resistance to the expanding rarefaction wave, is set by the density of residual gas. A suitable density enables the formation of a shock during the experimental observation window. The density of the residual gas and how much of it gets ionized cannot always be accurately determined, which complicates systematic parametric studies of interactions between rarefaction waves and ambient plasma in the lab. The early evolution of electrostatic shocks has been observed experimentally~\cite{Romagnani,Ahmed}.

Ablating two targets generates two rarefaction waves. A suitable target arrangement forces them to collide. If the flows are fast enough, the Weibel instability  between the counter-streaming ion beams is triggered, and collisionless shocks can form\cite{fiuza_weibel-instability-mediated_2012, fiuza_electron_2020}. Such fast flows require high-energy facilities. Here, we are interested in the interaction of slower flows, as in Refs. [3] and [4]. 

The most common configuration to create collisionless shocks is with two counter-streaming plasmas, as expected in most astrophysical environments, such as supernova ejecta interacting with the circumstellar medium. The interaction between counter-streaming plasmas to create electrostatic shocks has been studied theoretically without an ambient plasma\cite{Sorasio}. We also note that the collision of two electrostatic shocks is not well studied, but already exhibits interesting properties, such as the generation of confined magnetic fields\cite{boella_interaction_2022}. The possibility of similar interaction, also exists, during the interaction between coronal mass ejecta\cite{liu_interactions_2012} or when a supernova remnant collides with a stellar wind\cite{velazquez_supernova_2003}. A similar situation arises in laboratory astrophysics when two expanding plasma plumes interact in the presence of a residual background gas. The present study constitutes a new step, investigating the interaction of two similar unmagnetized expanding plasmas, with or without an ambient plasma.

Each rarefaction wave interacts with the ambient plasma and the other rarefaction wave. In what follows, the term separation vector denotes the line between the centers of both dense plasmas. The distance between the surfaces of both dense plasmas along the separation vector determines the collision speed of the rarefaction waves in the absence of ambient plasma; greater separation results in a faster collision. 

Rarefaction waves are modified by ambient plasma. An electrostatic shock, or a hybrid structure combining a shock for ambient ions with a double layer for rarefaction wave ions, can alter the velocity and density profiles prior to collision of both rarefaction waves, thereby influencing the structures that form at the collision location.

In this study, we investigate the interaction of two radial rarefaction waves and the effects of ambient plasma on their collision using two-dimensional particle-in-cell (PIC) simulations. We consider an unmagnetized collisionless plasma, in which the plasma dynamics is determined by collective electric fields. Our simulation setup is similar to that described in~\cite{Radial1,Radial2,Radial3} for unmagnetized plasmas, and in~\cite{Related1, Related2} for studies including magnetic field effects. We compare the interaction of rarefaction waves in vacuum with that in an ambient plasma with two different relative densities. Fully ionized nitrogen ions are the sole carriers of positive charge.

In our first simulation, two rarefaction waves interpenetrate in vacuum and form an overlap layer. An exponential density decrease of the rarefaction waves with increasing distance from the rarefaction wave front leads to a density minimum at the center of the overlap layer. Its density profile can thus not drive a thermoelectric field that slows the incoming rarefaction wave ions, which is required to create shocks. The ion beams of both rarefaction waves lead to the growth of ion acoustic waves in the overlap layer.  

In a second simulation, both rarefaction waves expand into an ambient plasma. The thermoelectric field, which causes the expansion of the rarefaction wave, also accelerates the ambient ions. Each rarefaction wave drives one hybrid structure. Both hybrid structures collide halfway along the separation vector and carry the accelerated ambient plasma to their collision boundary. The ions of the ambient plasma accumulate at this boundary and form an ion slab with a vanishing mean speed. All ambient ions were swept out of the interval between the dense plasma and this ion slab causing the rarefaction wave to expand freely reaching supersonic speeds before it hits the ion slab. In the rest frame of the rarefaction wave ions, the thermoelectric field of the ion slab moves at a speed larger than the ion acoustic speed. It is strong enough to drive a shock wave. This reverse shock moves in the direction of the rarefaction wave front. Once reverse shocks have developed on both sides of the ion slab, the inflowing rarefaction wave exchanges momentum with the reverse shock and no longer with the ion slab; the ambient ions disperse.

In a third simulation, we reduce the ambient plasma density. An ion slab also forms in that simulation, but its thermoelectric field is not strong enough to create reverse shocks. 

The paper is organized as follows. Section~\ref{background} outlines the principles of the PIC simulation method and our initial conditions. Section~\ref{simulationresults} presents the results of the three PIC simulations. Our findings are discussed in Section~\ref{discussion}.

\section{\label{background}Background}

We use the PIC simulation code \emph{Smilei} for our simulations, which resolve two spatial dimensions. The code represents the electric field $\mathbf{E}$, the magnetic field $\mathbf{B}$, and the macroscopic electric current $\mathbf{J}$ on a numerical grid. The electric current evolves the electromagnetic fields using Amp\`ere's law and Faraday's law. The magnetic divergence law and Gauss' law are fulfilled to round-off precision. The electrons and fully ionized nitrogen ions are resolved by an ensemble of computational particles (CPs), which have the same charge-to-mass ratio as the plasma particles they represent. We employ the correct ion-to-electron mass ratio $m_i / m_e \approx 2.6 \times 10^4$. A full description of the simulation code is given elsewhere~\cite{Smilei}.  

\begin{figure}[h]
    \includegraphics[width=0.7\columnwidth]{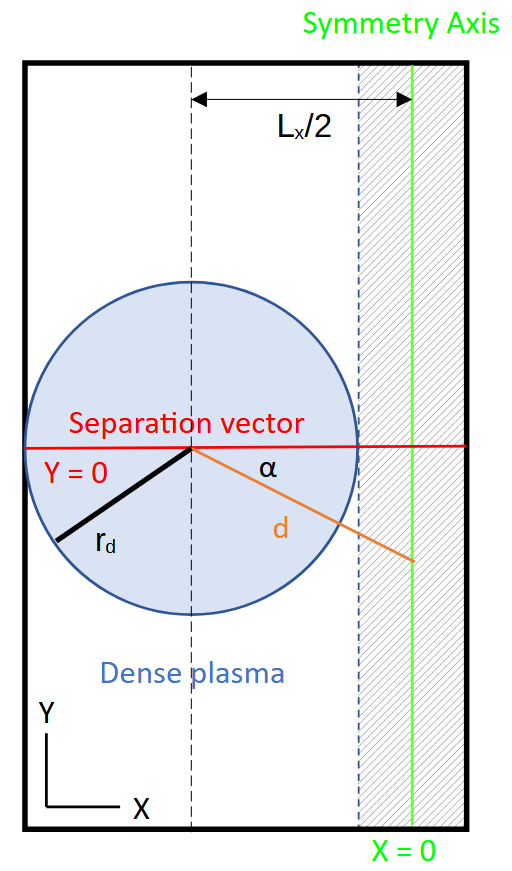}
    \caption{A circular dense plasma with radius $r_D$ is placed in a 2D simulation box, which resolves $- (L_x / 2 + r_D) \le x \le L_x/2 -r_D$ and $-L_y/2 \le y \le L_y/2$. The circular dense plasma is centered at $(x_c, y_c) =(-L_x/2,0)$. The separation vector starts at $(x_c,y_c)$ and returns to this point after crossing the right boundary. The vector $\mathbf{d}$ starts at $(x_c,y_c)$ and ends at the point on the symmetry axis $x=0$ that is defined by its angle $\alpha$ relative to the separation vector. The space outside the circle is either vacuum or ambient plasma with the same fully ionized nitrogen ions as the dense plasma. The hashed zone corresponds to the displayed zone in the following figures.}
    \label{InitialCondition}
\end{figure}

Figure~\ref{InitialCondition} sketches the initial plasma distribution in our simulations. The simulations with periodic boundary conditions resolve $L_x = 2.95$~mm along $x$ with 1024 grid cells and $L_y = 8.85$~mm along $y$ with 3072 grid cells.

The box resolves the intervals $-L_x/2-r_D \le x \le L_x / 2 -r_D$ and $-L_y / 2 \le y \le L_y/2$. The center of a circular dense plasma with constant density and radius $r_D=1.1$~mm is located at $(x_c,y_c)=(-L_x / 2, 0)$. Its electrons and nitrogen ions have the temperatures $T_{e0}=1500$~eV and $T_{i0}=100$~eV and number densities $n_{e0}=10^{16}$~cm$^{-3}$ and $n_{i0}=n_{e0}/7$. The electrons and ions of the dense plasma are resolved by 160 computational particles (CPs) per cell each. 

The electron plasma frequency and thermal speed are $\omega_{pe}=(n_{e0}e^2/\epsilon_0 m_e)^{1/2}$ and $v_{th,e}=(k_BT_{e0}/m_e)^{1/2}$
with the Boltzmann constant $k_B$ and elementary charge $e$. The values $\omega_{pe} = 5.64 \times 10^{12}$~s$^{-1}$ and $v_{th,e} = 1.6\times 10^7$~m/s give the Debye length $\lambda_D = v_{th,e}/\omega_{pe}\approx 2.9\times 10^{-6}$~m and $r_D=384\lambda_D$. The grid cell size is set equal to $\lambda_D$ and the timestep to 0.95$\lambda_D/c$.
The ion acoustic speed $c_s=(k_B(ZT_{e0}+3T_{i0})/m_i)^{1/2}$ with the ion charge $Z=7$ is $c_s \approx 3.5 \times 10^5$~m/s. The adiabatic constants are 1 for electrons and 3 for ions~\cite{Noise}.

At time $t=0$, the rarefaction wave front follows the boundary of the circle. It propagates towards the center of the circle $(x_c,y_c)$ at the speed $\approx c_s$. Rarefaction wave ions are accelerated in the radially outward direction; the rarefaction wave and the rarefaction wave front propagate in opposite directions. The expansion of the rarefaction wave across the vertical periodic boundary at $-L_x/2-r_D$ leads to its self-collision on the symmetry axis $x=0$. The larger vertical box size delays the self-collision along $y$. 

In simulation~1, the rarefaction wave expands into a vacuum. The slope of the velocity profile decreases in time because the ions propagate while they accelerate. The slope is invariant in dimensionless coordinates $\xi=x/(c_st)$, which assume a self-similar expansion~\cite{Gurevich66, crow_expansion_1975} and scales only with time $t$. Ions gain speed mostly at early times, and their peak speed is comparable to $10c_{s}$.

The length $d$ of the distance vector $\mathbf{d}$ in Fig.~\ref{InitialCondition} increases with its angle $\alpha$ relative to the separation vector. At time $t=0$, the distance of the symmetry axis from the dense plasma front is $d^*=L_x/(2\cos{\alpha})-r_D$. At a given time, the mean speed of the ions along $\mathbf{d}$ is $v_i\propto d^*$ near the symmetry axis $x=0$. This speed decreases with time. Its projection along $x$ is $v_i\cos{\alpha}$. 

Rarefaction waves arriving from both sides form an overlap layer near $x=0$, which changes the density distribution and, thus, the thermoelectric field near the symmetry axis. 

Ambient plasma is present in simulations~2 and~3. The ambient plasma, which is expelled by the rarefaction wave, modifies the density profile, and thus the amplitude of the thermoelectric field that accelerates the ions. A changing acceleration modifies the profile of the ion's mean velocity as a function of their distance from the rarefaction wave front. If the ambient plasma density is high enough, a shock emerges. The radial shock speed is $v_{i,max}$. It is approximately constant and below that of the rarefaction wave ions ahead of it. The shock velocity component along $x$ is $v_{i,max}\cos{\alpha}$. As $\alpha$ increases, the collision speed of the shocks along the separation vector near $x=0$ decreases and becomes subsonic for a large $\alpha$. 

In simulation~2, the rarefaction wave expands into an ambient plasma with densities of $n_{e0}/50$ and $n_{i0}/50$. Simulation~3 considers the values $n_{e0}/75$ and $n_{i0}/75$. At time $t=0$, the smallest distance $L_x-2r_D$ between the circle boundaries is 256$\lambda_D$. This distance corresponds to 30$\lambda^*_D$ in simulation~2 and 24$\lambda^*_D$ in simulation~3, where $\lambda_D^*$ is the Debye length of the ambient plasma.

\section{\label{simulationresults}Simulation results}

\subsection{Simulation 1: Free expansion}

The dense plasma is surrounded by vacuum at the simulation's start. It expands in the form of a circularly symmetric rarefaction wave. Ions with $x<x_c$, which expand to the left in Fig.~\ref{InitialCondition}, cross the periodic simulation boundary and collide with the ions that expand to the right. 

\begin{figure}[h]
    \centering
    \includegraphics[width=\columnwidth]{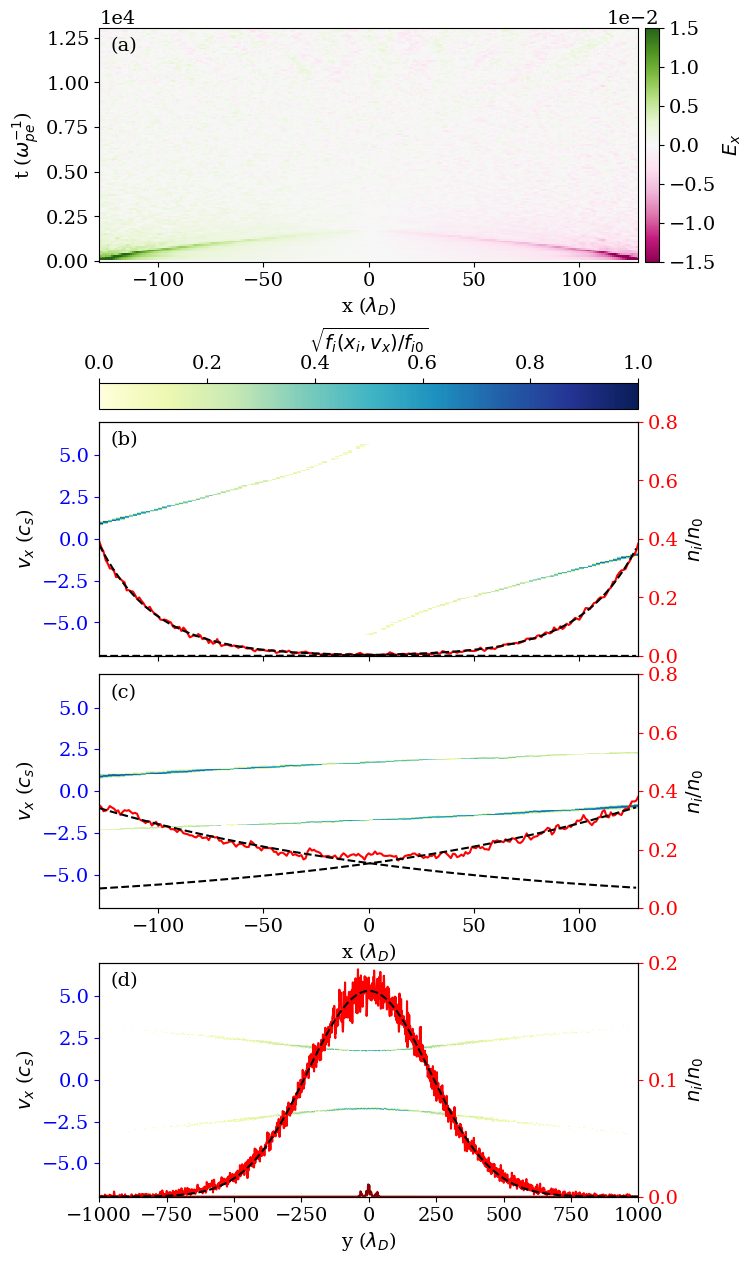}
    \caption{(a) x-component of the electric field, $eE_x/(m_ec\omega_{pe})$, along the separation vector. (b)-(d) Square root of the ion phase space densities $f_i(x_i, v_x)$, $x_i$ refers to x in panels (b,c), and y for panel (d), normalized to the peak value in the dense plasma at time t = 0, $f_{i0}$, and ion number densities of Simulations 1. Panels~(b) and~(c) show $f_i(x, v_x)^{1/2}$ and $n_i(x)$ along the separation vector at times $1800\omega_{pe}^{-1}$ and $7000 \omega_{pe}^{-1}$, respectively. The dashed curves are exponential fits. Panel~(d) shows ${f_i(y, v_x)}^{1/2}$ and $n_i(y)$ along the symmetry axis x = 0, at $7000 \omega_{pe}^{-1}$ (red curve) and $1800\omega_{pe}^{-1}$ (brown curve). The dashed curve is a Maxwellian fit. In all panels, the color scale and the velocity scale to the left apply to the phase space densities while the density scale to the right applies to the number densities.}
    \label{PS_sim1}
\end{figure}

Figure~\ref{PS_sim1}(a) shows the x-component of the electric field, $eE_x/(m_ec\omega_{pe})$, along the separation vector in function of position and time and Figure~\ref{PS_sim1}(b)-(d) shows the phase space density and number density of the ions at several times and along several directions.
Figure~\ref{PS_sim1}(b) shows both distributions along the separation vector $y=0$ at time $t=1800\omega_{pe}^{-1}$. The range of $x$ covers the vacuum gap at $t=0$. At the interval boundaries, the ion density is about $0.4 n_{i0}$. It reaches $n_{i0}$ at the rarefaction wave front, which is located outside of the displayed interval. Both rarefaction waves have a density profile that decreases exponentially with distance from the rarefaction wave front. 

The ions of both rarefaction waves have a mean velocity modulus $|\langle v_x \rangle|$ along $x$ that changes linearly with $x$ for $|x| > 45\lambda_D$ and faster than linear at smaller $x$. Self-similarity breaks down when acceleration takes place on scales comparable to $\lambda_D$~\cite{Gurevich66}, which was the case just after the simulation started. The value of $|\langle v_x \rangle|$ is $\approx c_s$ near $|x|\approx 130\lambda_D$ and increases to $5c_s$ near $x=0$. Their thermal spread is small. The ions were cooler than the electrons to start with and their large mass implies a low thermal speed. Their thermal spread also diminishes with time in a rarefaction wave~\cite{Gurevich66}.

Figure~\ref{PS_sim1}(d) shows both profiles at time $7000\omega_{pe}^{-1}$. The ion density $\approx n_{i0}/5$ is almost constant in the interval $-30\lambda_D \le x \le 30\lambda_D$, causing the thermoelectric field to vanish, and confirmed by Fig.~\ref{PS_sim1}(a). A weak electric field implies that the observed spatial change of $|\langle v_x \rangle|$ was established before the rarefaction waves overlapped. As shown in Fig.~\ref{PS_sim1}(a) before 1800$\omega_{pe}^{-1}$ the thermoelectic field scales with the inverse time and is strongest at early times~\cite{Gurevich66}. Rarefaction wave ions are decelerated along $x$ by the thermoelectric field of the density gradient on the opposite side of their rarefaction wave front. 

Figure~\ref{PS_sim1}(c) shows the density and phase space density distributions of the ions along the symmetry axis $x=0$. Fluctuations of the ion density near $y=0$ evidence that the rarefaction waves just reached the position $x=0$ at $t=1800\omega_{pe}^{-1}$. At time $t=7000\omega_{pe}^{-1}$, $n_i(y)$ follows a Maxwellian distribution. 

The number density of a rarefaction wave decreases exponentially with distance from its front. Consider a rarefaction wave front surrounding a point-like dense plasma located at $(x_c,y_c)$ with $x_c < 0$ and $y_c=0$. The distance 
\begin{equation}
    d=((x-x_c)^2+y^2)^{1/2}
\end{equation}
to a point on the symmetry axis is $d=(x_c^2+y^2)^{1/2}$. The contribution of the right-moving rarefaction wave to the density at $x=0$ is

\begin{equation}
    n_i(y) \propto \exp{(-x_c)}\cdot \exp{(-0.5\cdot y^2)}.
\end{equation}

The weak thermoelectric field near $x=0$ could not compress the ions and the total ion density is the sum of the densities of both rarefaction waves.

The density distribution near $x=0$ and $y=0$ is a saddle point. Its thermoelectric field accelerates the ions in the direction of increasing $|y|$. Variations in the ion density of $0.1n_{i0}$ over the distance 250$\lambda_D$ in Fig.~\ref{PS_sim1}(d) yield an acceleration that is weak even compared to the barely visible slowdown of the ions by the ion density ramps in Fig.~\ref{PS_sim1}(c). The value of $|\langle v_x \rangle|$ along $y$ of the ions should thus be close to that expected for free expansion. Figure~\ref{PS_sim1}(d) shows that near the symmetry axis $x=0$ the value of $|\langle v_x \rangle|$ increases with $|\alpha|$. Its variation is $c_s/2$ between $y=0$ or $\alpha =0$ and $y=500\lambda_D$ or $\alpha \approx 45^\circ$. 

Several instabilities can develop in rarefaction waves and in the overlap layer. Rarefaction wave ions speed up and electrons slow as they move in the radially outward direction. Most electrons are reflected by the thermoelectric field. This reflection is quasi-elastic because the rarefaction wave is expanding\cite{mora_self-similar_1979}. Electrons lose momentum in the radial direction. They develop a thermal pressure anisotropy, which leads to the growth of a magnetic field. In our simulations, the magnetic field $\mathbf{B}$ driven by this instability remains weak. The peak electron gyro frequency $\omega_{ce}=e|\mathbf{B}|/m_e$ reaches only $\omega_{ce}/\omega_{pe}\approx 0.01$. We neglect this weak magnetic field.  

Ion beams become unstable if they counter-stream at a velocity that is large compared to their thermal speed. What type of instability dominates depends on how the relative speed of the ion beams compares to the electron's thermal speed and how uniform the plasma is. 

Consider electrons with a Maxwellian velocity distribution with a thermal spread close to $v_{th,e}$. Their thermal speed $v_{th,e} \approx 46c_s$ implies that the relative speed of the ion beams is small. Small and symmetric variations of the ion density and beam distributions around $x=0$ imply a low net electric current, which favors the resonant ion-ion instability over the current-driven ion acoustic instability. 

\begin{figure}[h]
    \centering
    \includegraphics[width=1\columnwidth]{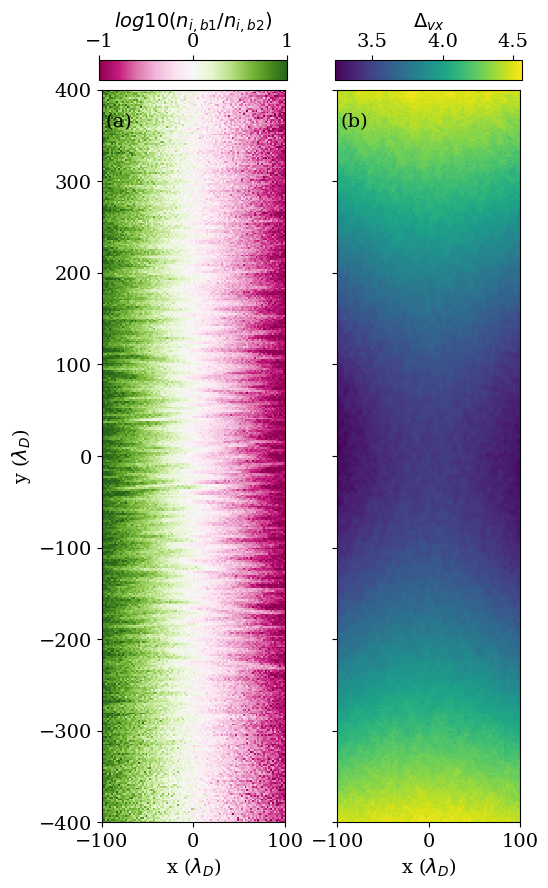}
    \caption{Panel~(a) shows the ratio $log10(n_{i,b1}/n_{i,b2})$ of the density $n_{i,b1}$ of beam~1 over $n_{i,b2}$ of beam~2. Panel~(b) shows the difference $\Delta_{vx}=(\langle v_x \rangle_{b1}+\langle v_x \rangle_{b2})/c_s$ of the mean velocities along $x$ of both beams. Both panels show the values at time $7000\omega_{pe}^{-1}$.}
    \label{Ratio}
\end{figure}

\begin{figure*}
    \centering
    \includegraphics[width=1\textwidth]{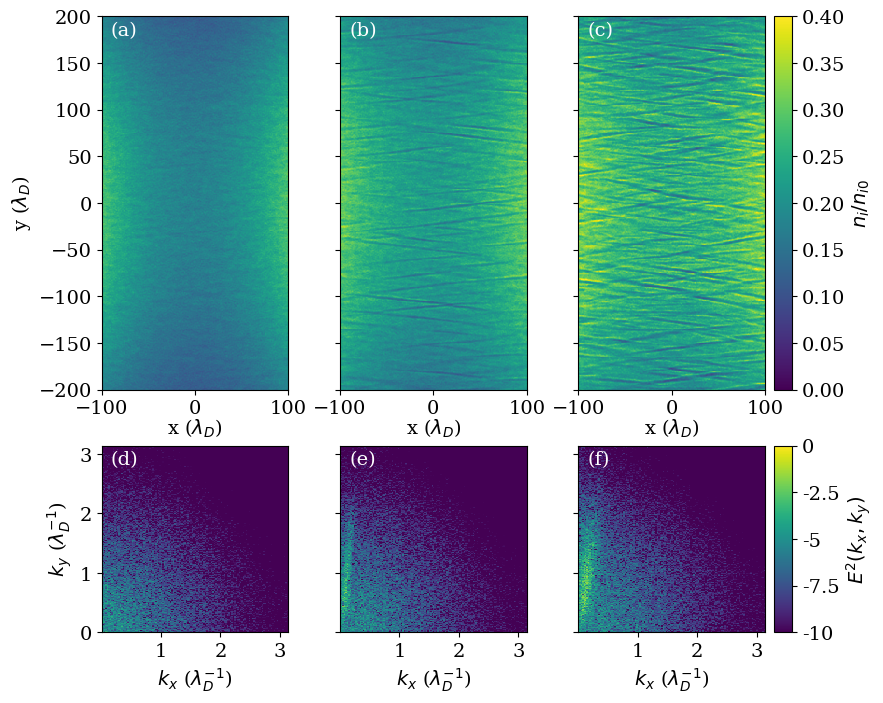}
    \caption{On the upper line, ion density $n_i/n_{i0}$ of Simulation 1 at times (a) $7000\omega_{pe}^{-1}$, (b) $8600 \omega_{pe}^{-1}$, and (c) $10200\omega_{pe}^{-1}$. On the bottom line power spectrum of the electric field $E_x^2(k_x, k_y) + E_y^2(k_x, k_y)$ normalized to its maximum value at 7000$\omega_{pe}^{-1}$ (panel (d)) corresponding to the above region at the same time. All panels on the same line share the same color scale.}
    \label{Dens_sim1}
\end{figure*}

The wavenumber spectrum and growth rate of the ion-ion instability depend on the relative velocity between the ion beams and their density ratio. Equal densities of the ion beams result in the aperiodic growth of waves\cite{stringer_electrostatic_1964}. Such waves do not propagate in the box frame and remain close to $x=0$ where the plasma is unstable. Beam speeds in excess of $c_s$ rotate their wavevector relative to the beam velocity vector.

Beam~1 and~2 denote the beams formed by the rarefaction wave ions that arrive at a point in the x-y plane from lower and larger $x$, respectively. Near $x=0$, both beams have equal speeds along $y$ and counter-stream along $x$.

Figure~\ref{Ratio}(a) shows the ratio of the density of beam~1 to that of beam~2 at $7000\omega_{pe}^{-1}$. Beam~1 is 5 times denser than beam~2 near $x=-100\lambda_D$ while beam~2 is denser than beam~1 by the same factor near $x=100\lambda_D$. Both beams have densities that differ by the factor 2 in the interval $-30\lambda_D \le x \le 30\lambda_D$. Symmetry implies that their ratio does not change much along $y$.  

Figure~\ref{Ratio}(b) shows the difference $\Delta_{vx}=(\langle v_x \rangle_{b1}+\langle v_x \rangle_{b2})/c_s$ of the mean velocities of beam~1 and~2. The value of $c_s$ is the ion acoustic speed in the dense plasma at $t=0$. Close to the symmetry axis $x=0$, $\Delta_{vx}$ has its minimum $3.4c_s$ at $y\approx 0$. It increases to 4.6$c_s$ near $|y| = 400\lambda_D$. Moderate changes in $\omega_{pe}\propto n_e^{1/2}$ and $\Delta_{vx}$ over extended spatial intervals along $x$ and $y$, together with $\Delta_{vx}\ge 1$, lead to ion-ion instability. 

Figure~\ref{Dens_sim1}(a) shows the ion density at time 7000$\omega_{pe}^{-1}$ and figure~\ref{Dens_sim1}(d) shows the spatial spectrum of the electric field, $E = E_x^2(k_x, k_y) + E_y^2(k_x, k_y)$, at the same time and for the same region. All the spatial spectra in Fig.~\ref{Dens_sim1} are normalized to the peak power in Fig.~\ref{Dens_sim1}(d). At this time, no particular wavenumber emerges from the noise.

Weak oblique striations are visible, at time $8600 \omega_{pe}^{-1}$ in Fig.~\ref{Dens_sim1}(b), the quasi-planar striations have grown to a large amplitude and they reached their peak amplitude in Fig.~\ref{Dens_sim1}(c). The length and thickness of the striations are of the order $100\lambda_D$ and $3\lambda_D$, respectively. It can also be seen in Fig.~\ref{Dens_sim1}(e) and (f), waves at low $k_x$ have gained power over the $k_y$ interval 0 to 2. The striations propagate slowly in the simulation frame. They form angles $\theta$ with x-axis with a modulus that ranges between $15^\circ$ and $17^\circ$. 

Oblique quasi-planar striations are a signature of ion-ion instabilities\cite{forslund_numerical_1970, gary_ion-ion_1987}. Their growth rate is comparable to the plasma frequency of ions $\omega_{pi}\approx \omega_{pe}/60$ for fully ionized nitrogen. Since the ion-ion instability only grows if both beams overlap and if their velocity difference does not change too much in space and time, the waves just started to grow at time 7000$\omega_{pe}^{-1}$. The time intervals 1600$\omega_{pe}^{-1}$ between Figs.~\ref{Dens_sim1}(a,~b) and between Figs.~\ref{Dens_sim1}(b,~c) equal $26\omega_{pi}^{-1}$, which is reduced by a factor of about 2 near the symmetry axis due to the low ion density $\approx 0.2n_{i0}$. The waves, which started to grow in Fig.~\ref{Dens_sim1}(a), had a few e-foldings to grow before saturating in Fig.~\ref{Dens_sim1}(c). The maximum spatial spectrum in Fig.~\ref{Dens_sim1}(e)-(f) follows the growth rate of the ion-ion acoustic instability in Fig~2 in Ref[23].

Ion acoustic waves, which are driven by two ion beams with relative speed $c_s$, have a wave vector that is parallel to their drift velocity. In this case, the striations would align with $y$. If the drift speed along $x$ exceeds $c_s$, the wave vector changes direction relative to the drift velocity vector\cite{chapman_nonlinear_2021}. This rotation reduces the projection of the beam velocity on the wavevector to $\approx c_s$. The angle of the striation wavevector relative to the separation vector near $y=0$ is $\approx 74^\circ$. Figure~\ref{Ratio} gives the value $\Delta_{vx}\approx 3.4c_s$. Its projection on the wavevector gives the speed $3.4c_s \cdot \cos{74^\circ} \approx 0.94c_s$. 

Simulation~1 showed freely expanding and overlapping rarefaction waves. The constant ion density in the overlap layer implied that there was no jump in the electric potential between the overlap layer and the inflowing rarefaction wave ions. They propagated through the overlap layer and caused the growth of ion acoustic waves. This holds for any collision between freely expanding rarefaction waves. Consider the two-dimensional density distribution with $x_0>0$  
\begin{equation}
    n_i(x,y) = \exp{(-r_1(x,y))} + \exp{(-r_2(x,y))}.
\end{equation}

with $r_1={((x+x_0)^2+y^2)}^{1/2}$ and $r_2={((x-x_0)^2+y^2)}^{1/2}$. By looking at its partial derivative along x, at $x=0$, we obtain $r_1(0,y)=r_2(0,y)\equiv R$ with $R>0$:

\begin{equation}
    \frac{\partial n_i}{\partial x}|_{x=0} = -\exp(R) \left(\frac{x_0}{R} - \frac{x_0}{R}\right) = 0.
\end{equation}

With the derivative of $n_i$ negative when x approaches 0 for x<0 and positive when x approaches 0 for x>0, the ion density $n_i(x,y)$ on the symmetry axis $x=0$ is always a minimum. Ion heating may eventually compress the ions near $x=0$ but this process is too slow in our simulations.

%In the interaction region we are in the presence of two counter-streaming of ions beams with almost the same density and with an electrons population we can consider at rest (since $v_{the} \gg cs$) as a neutralizing background. This configuration trigger the ion-ion acoustic instability. The electron-ion acoustic instability can only grows with a net current. In the rest frame of this system the instability has a zero real frequency\cite{stringer_electrostatic_1964}. So the waves generate by the ion-ion acoustic instability do not propagate in the plasma at rest. The plasma is not at rest, it has a net mean velocity in y due to the radial expansion. There is no net mean velocity in the x-direction since the beams have almost the same mean velocity but in opposite direction.
%The difference of mean velocity of beams in the x-direction is supersonic. This high counter-streaming velocity creates waves with a wave number \textbf{k} which has a large angle with the propagation direction of the beams. The higher the difference of velocity the larger the angle is\cite{chapman_nonlinear_2021}. As we moves away from the separation vector the difference in velocity increases (Fig.~\ref{Ratio}(b)). This explain why the angle between the waves (perpendicular to \textbf{k}) is smaller at large y. Compare to simulation with constant density profile (Fig.4 in Ref\cite{chapman_nonlinear_2021}) there are less waves growing. The non-uniform density does not prevent the growing of the ion-ion acoustic instability and the formation of waves but there less waves formed.

\subsection{Simulation 2: Expansion in dense ambient plasma}

%- Formation of a shock in Simulation 2 but not in Simulation 3 at the separation vector

%- In Simulation 3 shocks form away from the separation vector

%- The shock appears at the same time for ~200 lambda_D

%- In Simulation 2 the shock covers all the interaction region in along y

The dense plasma is surrounded by an ambient plasma with the same composition and temperature and a density that is reduced by the factor 50. 

\begin{figure}
    \centering
    \includegraphics[width=\columnwidth]{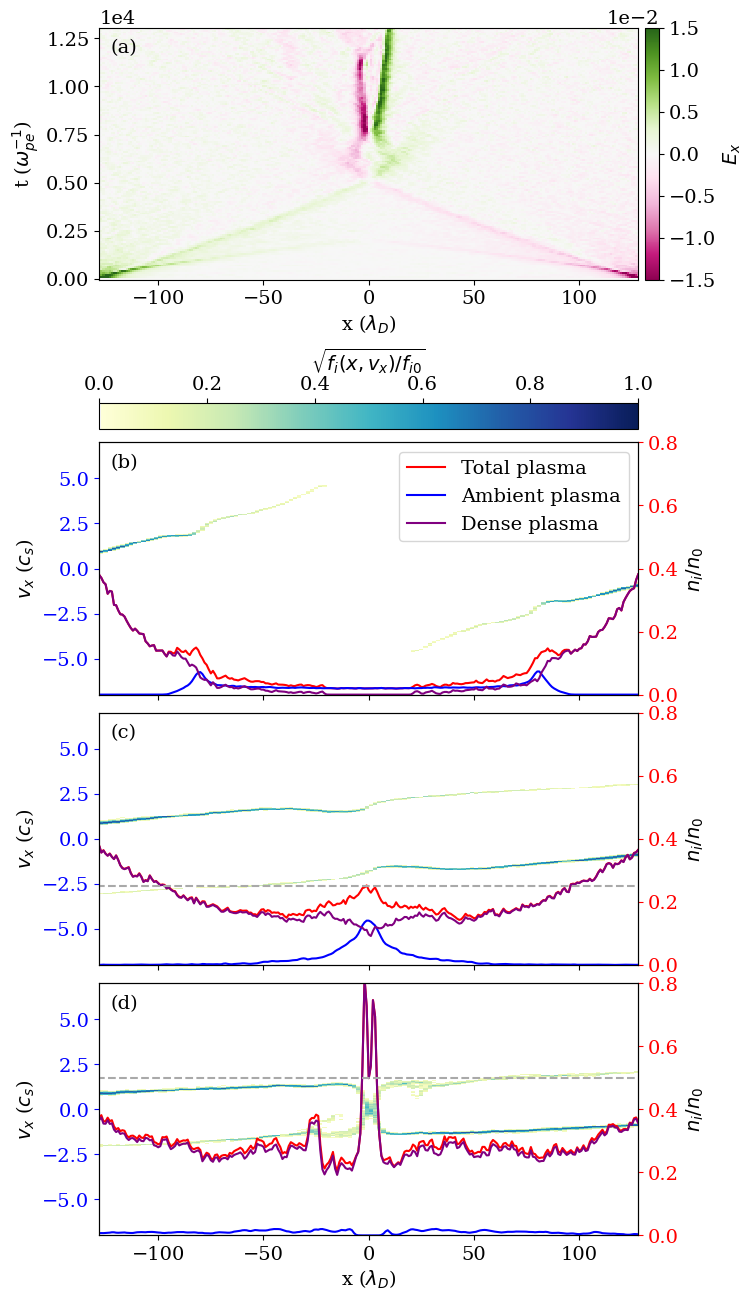}
    \caption{(a) x-component of the electric field $E_x$ along the separation vector. (b)-(d) Square root of the dense ion phase space densities $f_i(x, v_x)$, normalized to the peak value in the dense plasma at time t=0, $f_{i0}$, and ion densities of the dense plasma, the ambient plasma, and their sum of Simulation 2. Panel~(b) shows $f_i(x,v_x)^{1/2}$ and the ion densities at time $1800\omega_{pe}^{-1}$ and along the separation vector. Panels~(c) and (d) show the same quantities at the times $5200\omega_{pe}^{-1}$ and $8600\omega_{pe}^{-1}$, respectively. The velocity scale to the left and the color map apply to $f_i(x,v_x)$. The density scale to the right and the inset apply to the ion densities. The grey dashed line highlight the density of the hybrid structure in (c) and the reverse shock density in (d).
    }    
    %\caption{Square root of the dense ions phase space distribution for the blue map and ions density for the curves, in yellow the density of the ambient plasma, in purple the density of the dense plasma and in red the total density, of Simulation 2, at the separation vector at different time. (a) $1800\omega_{pe}^{-1}$. (b) $5200 \omega_{pe}^{-1}$. (c) $8600 \omega_{pe}^{-1}$.}
    \label{PS_sim2}
\end{figure}

Figure~\ref{PS_sim2}(a) shows the x-component of the electric field along the separation vector. The first signal, which disappears around $2000\omega_{pe}^{-1}$, corresponds to the nearly free expansion of the rarefaction wave, similarly to Fig.~\ref{PS_sim1}(a). 
Figure~\ref{PS_sim2}(b) shows the ion density and the ion phase space density at time $1800\omega_{pe}$. 
The ions of the ambient plasma have been pushed out of the interval $|x|>100\lambda_D$ and the rarefaction wave is freely expanding. Like in the case of simulation~1, the ion beam has a value $|\langle v_x \rangle| $ that is proportional to its distance from the rarefaction wave front and an exponential decrease in ion density. Ambient ions have been piled up in the interval $85\lambda_D \le |x| \le 100 \lambda_D$. Their number density is comparable to that of the rarefaction wave ions, and the total ion density in this interval is constant. The constant value of $|\langle v_x \rangle |$ in this interval is typical for the velocity component along the shock density gradient of the ions downstream of a hybrid structure. 

A jump in the ion density at the front of the density plateau is the transition layer of an electrostatic shock. Its thermoelectric field, visible in (a) between 0 and 5200$\omega_{pe}^{-1}$, accelerates the ambient ions at rest in the direction of $x=0$. The ambient ions that cross the transition layer enter its downstream plasma. Ambient ions, which did not overcome the electric field, form the shock-reflected ion beam. Rarefaction wave ions, which catch up with the electrostatic shock, are accelerated by the electric field as can be seen from the ion phase space density distribution. The electrostatic shock therefore acts as a double layer for the rarefaction wave ions. Fast ions in the interval $|x| < 50 \lambda_D$ were accelerated by the rarefaction wave prior to the formation of the hybrid structure. They have reached $|x| \approx 20\lambda_D$ at time $1800\omega_{pe}^{-1}$.

Figure~\ref{PS_sim2}(c) shows the same quantities at time $5200\omega_{pe}^{-1}$. Ambient ions are only found in the interval $|x|<50\lambda_D$. Their density increases steadily from the interval boundary until $x=0$, where they reach $0.14n_{i0}$. This density is several times that found at the earlier time near $|x|=80\lambda_D$. The total ion density has a maximum near $x=0$. It decreases with increasing $|x|$ until it reaches the interval with the freely expanding rarefaction wave. After 5200$\omega_{pe}^{-1}$ in Fig.~\ref{PS_sim2}(a) the sign of the thermoelectric field changed, due to the accumulation of ambient ions. Inflowing rarefaction wave ions are slowed and compressed near $|x|=20\lambda_D$ by the positive electric potential associated with the accumulation of the total ion density. The accumulation of ions from the rarefaction wave near $|x|=20\lambda_D$ reduces the number of ions that flow into the interval $|x| < 20\lambda_D$. 

\begin{figure}[h]
    \centering
    \includegraphics[width=\linewidth]{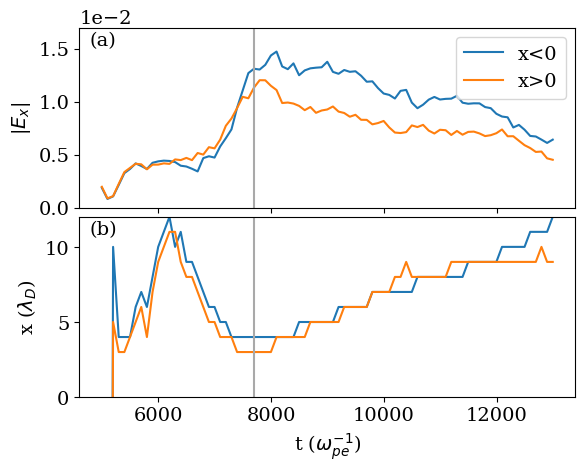}
    \caption{(a) Maximum value of $E_x$, average between y = -30$\lambda_D$ and 30$\lambda_D$, for the negative x in blue and positive x in orange. (b) Position of the maximum electric field, same legend as (a).}
    \label{Efield}
\end{figure}

Figure~\ref{Efield}(a) shows the maximum value of $E_x$, averaged over $y = -30\lambda_D$ to $30\lambda_D$, as a function of time after $5200\omega_{pe}^{-1}$. The blue curve corresponds to negative $x$, and the orange curve to positive $x$. Figure~\ref{Efield}(b) shows the position of the maximum value, using the same legend.
The electric field grows until $5700\omega_{pe}^{-1}$, where it reaches a first plateau. For about $800\omega_{pe}^{-1}$, the field remains constant, while the positions of the maxima move away from $x = 0$. The incoming ions from the rarefaction wave are slowed down and broaden the region of accumulated ions. At $6500\omega_{pe}^{-1}$, the electric field increases as the accumulation region narrows. The electric field reaches its maximum at $7700\omega_{pe}^{-1}$, indicated by a vertical gray line in Fig.~\ref{Efield}. The shocks are fully formed at this time.

In the rest frame of each ion beam, the accumulation of ambient ions corresponds to a fast-moving electric field. Its amplitude is large enough to create a density wave in the rarefaction wave ions, and its speed is large enough to turn the wave into an electrostatic shock. This shock flows in the direction of the rarefaction wave front. Thus, it is a reverse shock. In the laboratory frame the reverse shocks move at a constant speed (Fig.~\ref{Efield}(b)) of 20 km/s.

Figure~\ref{PS_sim2}(c) shows the pair of reverse shocks formed at time $8600\omega_{pe}^{-1}$. They are located at $|x|\approx 4\lambda_D$, and they reflect some but not all rarefaction wave ions that arrive from their respective upstream directions. Symmetry implies that the mean speed of the ions is zero downstream of both reverse shocks. The phase space density distribution of the ions in the downstream region $|x| < 10\lambda_D$ is not a single Maxwellian. The shocks did not reduce the mean speed of the ions to zero in the simulation box frame after they crossed the transition layer. The speed difference gave rise to vorticity creating a vortex in the ion phase space distribution. 

\begin{figure}[h]
    \centering
    \includegraphics[width=\linewidth]{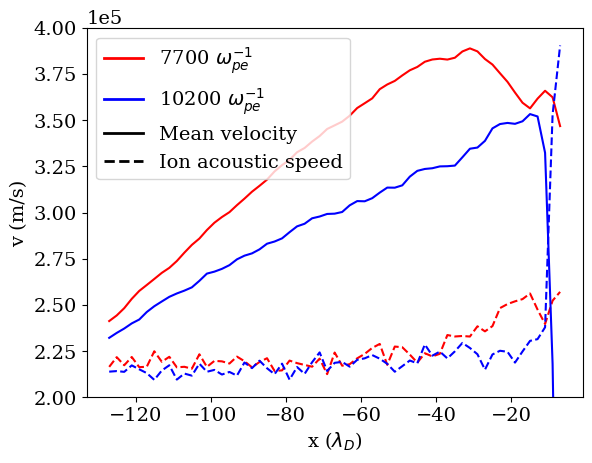}
    \caption{Mean velocity of the ion beam coming from the left in full line and ion acoustic speed in dashed line at two different time, 7700$\omega_{pe}^{-1}$ in red and 10200$\omega_{pe}^{-1}$ in blue.}
    \label{velocity}
\end{figure}

Determining the shock Mach number requires taking into account the mean ion velocity within the rarefaction wave. The mean velocity of the ions coming from the left is shown as a solid line in Figure~\ref{velocity}. At $7700\omega_{pe}^{-1}$ (red curve), between $x = -40$ and $-20\lambda_D$, the mean velocity decreases while the ion acoustic speed (dashed curve) increases. This is due to the thermoelectric field of the shock during its formation. We consider the unperturbed ions at $x = -40\lambda_D$ as the incoming ions. At the time of formation, the shock velocity is about $3.9 \times 10^5$ m/s. The dashed line in Fig.~\ref{velocity} represents the ion acoustic speed. The Mach number at $7700\omega_{pe}^{-1}$ is 1.7.
At a later time, the velocity of the incoming rarefaction wave is about $3.5 \times 10^5$ m/s at $x = 15\lambda_D$, and the Mach number is 1.6. As the velocity of the incoming ions decreases, the Mach number decreases accordingly.

\begin{figure*}
    \centering
    \includegraphics[width=\textwidth]{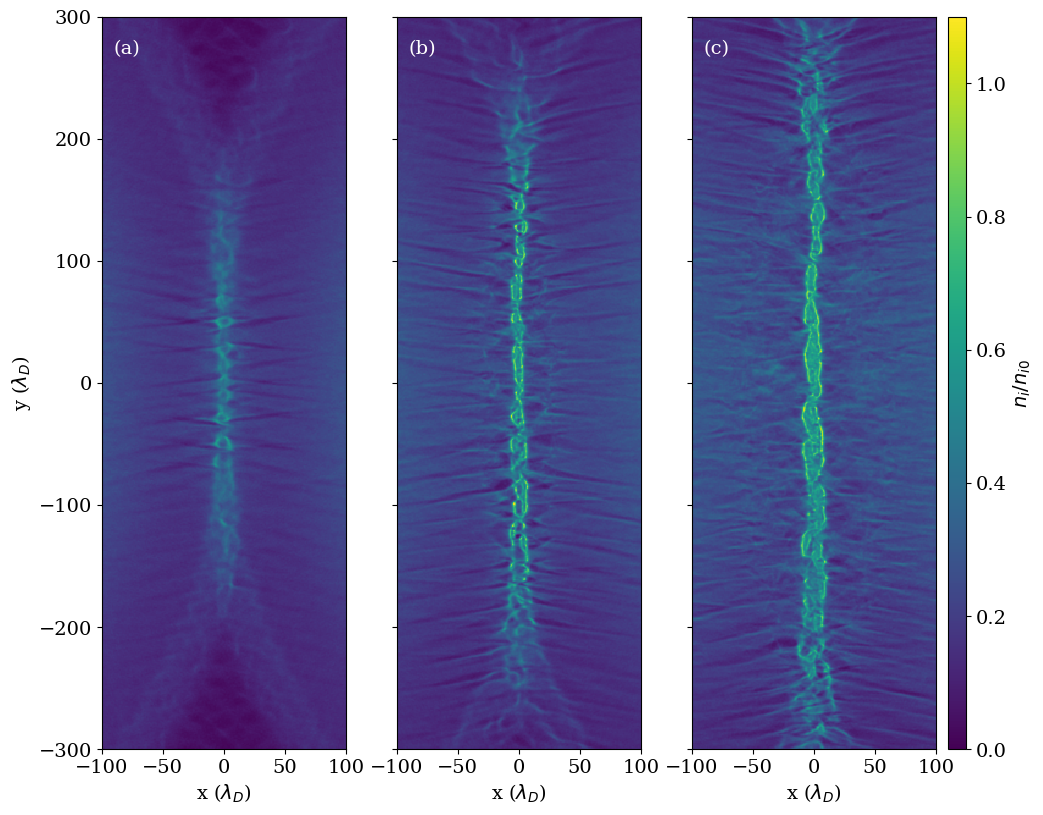}
    \caption{Ion density $n_i/n_{i0}$ of Simulation 2 at times (a) $7000\omega_{pe}^{-1}$, (b) $8600 \omega_{pe}^{-1}$, and (c) $10200\omega_{pe}^{-1}$. All panels share the same color scale. The vertical gray line}
    \label{Dens_sim2}
\end{figure*}

As long as the ion acoustic waves were growing in the inflowing rarefaction waves, the ions of the rarefaction wave transferred momentum to the compressed ambient ions, which served as a piston. This momentum transfer balanced the thermal pressure of the compressed and hot ambient ions. Once the reverse shocks detached from the piston, the momentum transfer ceased and the accumulation of ambient ions dispersed. Some ambient ions crossed the shocks, which acted as double layers for them.

The forward shock that expands radially outward from the point $x=x_c$ and $y=y_c$ and propagates in ambient plasma, is here an hybrid structure with a shock in the ambient plasma and a double layer for the dense plasma. A reverse shock propagates in the rarefaction wave plasma. According to Fig.~\ref{PS_sim2}(c), ion densities $\sim 0.5n_{i0}$ indicate that reverse shocks are present, while densities $\sim 0.3n_{i0}$ are representative of hybrid structures. 

Figure~\ref{Dens_sim2}(a) shows the total ion density at time $7000\omega_{pe}^{-1}$. Planar reverse shocks confine the downstream plasma for $|y|< 100\lambda_D$. At time $t=0$, the distance of the rarefaction wave front from $x=0$ is $128\lambda_D$. The distance increases in time as a result of the inward propagation of the rarefaction wave front. The values $|y| \le 100\lambda_D$ lead to a moderate increase in this distance. The arrival times of hybrid structures near $x=0$ should not differ too much in this $y$ interval and reverse shocks do not propagate rapidly in the box frame. The reverse shocks, which confine their shared downstream region, are thus approximately planar. In this $y$ interval, the ion density behind the reverse shocks is below that of the shocks. This density depression is characteristic of ion phase-space vortices\cite{eliasson_formation_2006}.

The interval with the enhanced ion density broadens for $100\lambda_D < |y| < 130\lambda_D$ at 7000$\omega_{pe}^{-1}$. In this interval, ambient ions are still being compressed by the converging rarefaction waves. This case corresponds to that depicted in Fig.~\ref{PS_sim2}(c). Ambient ions accumulate near $x=0$ only up to $|y|\approx 200\lambda_D$. Radial structures indicative of hybrid structures, which have not yet collided, are present for larger $y$. 

Figure~\ref{Dens_sim2}(b) shows a planar ion density structure near $x=0$ for $|y| <170\lambda_D$. Its density is uniform on scales $10\lambda_D$ along $x$ and close to that of the reverse shocks. A broader structure is located near $|y|=210\lambda_D$ and separate hybrid structures can be seen near $|y|=300\lambda_D$. The broader structure near $|y|=120\lambda_D$ at the earlier time has now changed into reverse shocks that enclose a shared downstream plasma. This corroborates that the broader structure corresponds to ambient ions, which are still being compressed. 

Figure~\ref{Dens_sim2}(c) shows an ion density near $x=0$ that is approximately uniform for $|y| < 170\lambda_D$. It becomes patchy with larger $y$. At this time, the hybrid structures gave way to reverse shocks, which are smooth only for $|y| < 150\lambda_D$. 

According to Fig.~\ref{PS_sim2}, reverse shocks start to grow along the separation vector at $5200\omega_{pe}^{-1}$ and have fully developed at $8600\omega_{pe}^{-1}$. During this time, the ambient ions are concentrated near the symmetry axis. Rarefaction waves expand freely up to the reverse shock. 

Figure~\ref{Ratio}(b) shows that the speed of the ions of the rarefaction wave projected on the separation vector, $\Delta_{vx}/2$, ranges from 1.7$c_s$ near $y=0$ to 2$c_s$ at $|y|=200\lambda_D$. A parametric study, which examined how the speed of an electrostatic piston in an electron-ion plasma affects the formation of a hybrid structure, was conducted in~\cite{Dieckmann2013}. Hybrid structures formed rapidly for speeds below $2.2c_s$ and with a delay for speeds up to $2.5c_s$. Changes in the smoothness of the ion density near $x=0$ may be related to the change from a laminar to unsteady electrostatic shock. Any non-uniformity of the ion slab is probably further amplified by the flow component of the rarefaction wave along $y$.

\subsection{Simulation 3: Expansion in a lower density ambient plasma}

The compressed ambient plasma led to the growth of reverse shocks. By reducing the density of the ambient plasma, we decrease the number of ambient ions that can be piled up by the converging rarefaction waves.  

\begin{figure}
    \centering
    \includegraphics[width=\columnwidth]{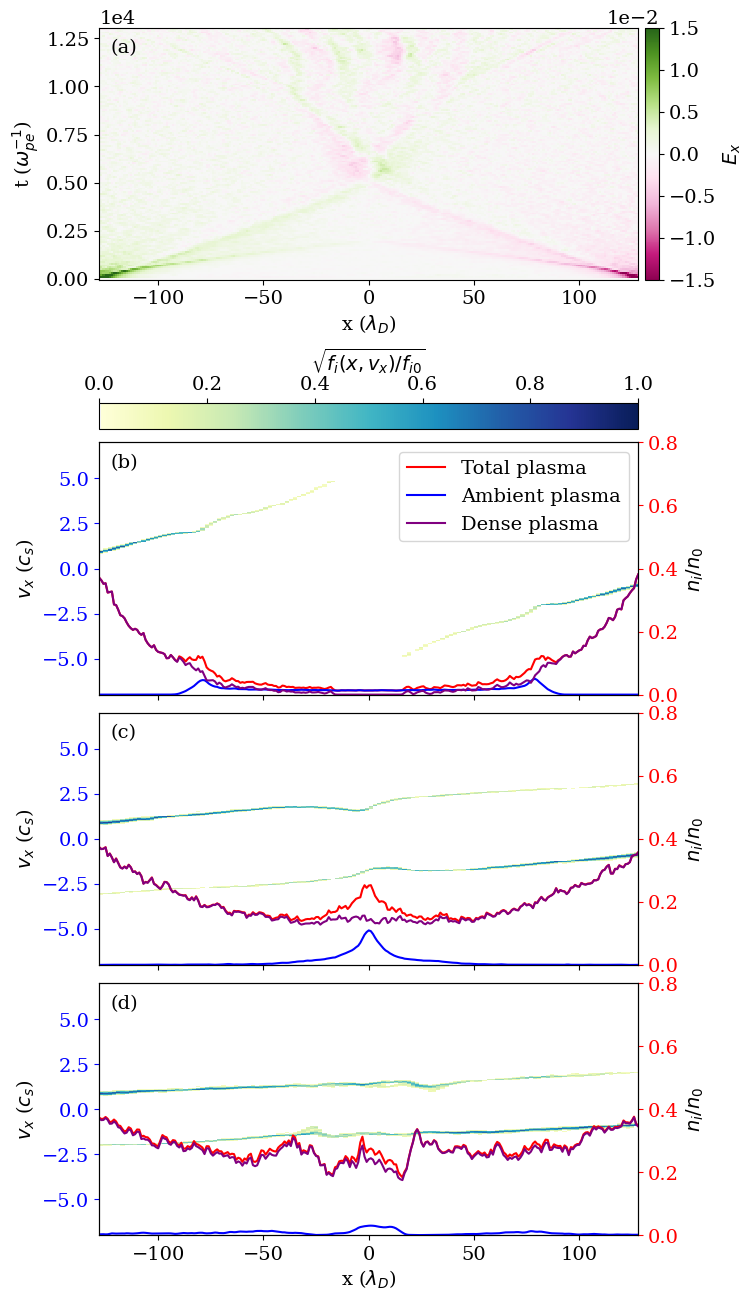}
    \caption{(a) x-component of the electric field $E_x$ along the separation vector. (b)-(d) Square root of the dense ion phase space densities $f_i(x, v_x)$, normalized to the peak value in the dense plasma at time t=0, $f_{i0}$ and ion densities of the dense plasma, the ambient plasma, and their sum of Simulation 3. Panel~(a) shows $f_i(x,v_x)^{1/2}$, and the ion densities at time $1800\omega_{pe}^{-1}$ and along the separation vector. Panels~(b) and (c) show the same quantities at the times $5200\omega_{pe}^{-1}$ and $8600\omega_{pe}^{-1}$, respectively. The velocity scale to the left and the color map apply to $f_i(x,v_x)$. The density scale to the right and the inset apply to the ion densities.     
    }
    %Square root of the ions phase space distribution for the blue map and ions density for the curves, in yellow the density of the ambient plasma, in purple the density of the dense plasma and in red the total density, of Simulation 3, at the separation vector at different time. (a) $1800\omega_{pe}^{-1}$. (b) $5200 \omega_{pe}^{-1}$. (c) $8600 \omega_{pe}^{-1}$.}
    \label{PS_sim3}
\end{figure}

Figure~\ref{PS_sim3}(b)-(d) shows the number densities and phase space densities of ions along the separation vector at the same times as Fig.~\ref{PS_sim2}(b)-(d). 
A comparison of the data at time $1800\omega_{pe}^{-1}$ shows that the ambient ions in Fig.~\ref{PS_sim3}(b) have been evacuated from a larger spatial interval than in simulation~2. Their number density goes to zero near $|x| = 90\lambda_D$ rather than $100\lambda_D$ as in simulation~2. The total ion density does not show the plateau, which is characteristic of the downstream region of a fully developed hybrid structure or electrostatic shock. 

The accumulated ambient ions decreased the change in the total ion density in the interval $80 < |x| < 90$, reducing the amplitude of the thermoelectric field, as represented in Fig.~\ref{PS_sim3}(a). Rarefaction wave ions do therefore not gain much speed in this interval. 

\begin{figure}
    \centering
    \includegraphics[width=\columnwidth]{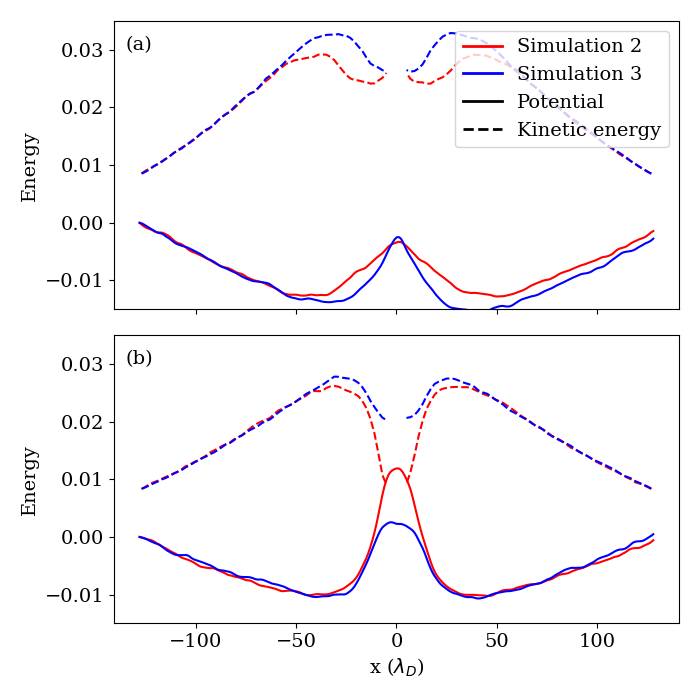}
    \caption{Potential of $E_x$ average between y = $\pm$ 30$\lambda_D$ in dashed line and the kinetic energy of the ion beams in full line for Simulation 2 (in red) and 3 (in blue) at (a) 5200 $\omega_{pe}^{-1}$ and (b) 6500$\omega_{pe}^{-1}$.}
    \label{energy}
\end{figure}

At time $5200\omega_{pe}^{-1}$, the ambient ions are concentrated near $x=0$ forming an electrostatic piston that moves at a speed well above $c_s$ in the local rest frame of the rarefaction wave ions. Figure~\ref{energy}(a) shows the potential, averaged over $y = \pm 30\lambda_D$ (solid line), and the kinetic energy (dashed line) for simulations~2 (red) and 3 (blue). The peak potential amplitude is similar in both cases, but it is broader in simulation~2. In this case, the thermoelectric field slows down the ions of the rarefaction waves for a longer time. In simulation~3, the thermoelectric field is strong enough to slow the ions of each rarefaction wave as seen from their phase-space density distribution in Fig.~\ref{PS_sim3}(c). However, this slowdown is not strong enough to result in a density change of the rarefaction wave ions, as shown in Fig.~\ref{energy}(b).

Such a density change would add to the number of ions concentrated near $x=0$ and thus to the amplitude of its positive electric potential relative to the surrounding rarefaction waves. A larger electric potential difference slows the rarefaction wave ions even more. This feedback loop led to reverse shocks in simulation~2. The thin ambient plasma in simulation~3 can apparently not initiate this feedback loop.

Figure~\ref{PS_sim3}(c) confirms that no reverse shocks formed in simulation~3. Velocity oscillations in the ion beam~1 with $v_x > 0$ are present near $x=40\lambda_D$. Ion beam~2 has velocity oscillations near $x=-25\lambda_D$. Velocity oscillations cause strong oscillations of the ion density and trap some ambient ions. A comparison of Figs.~\ref{PS_sim3}(b,~c) shows that these oscillations correspond to rarefaction wave ions that were slowed down by the piston and convected with the rarefaction wave once the ambient ions dispersed.

\begin{figure*}
    \centering
    \includegraphics[width=\textwidth]{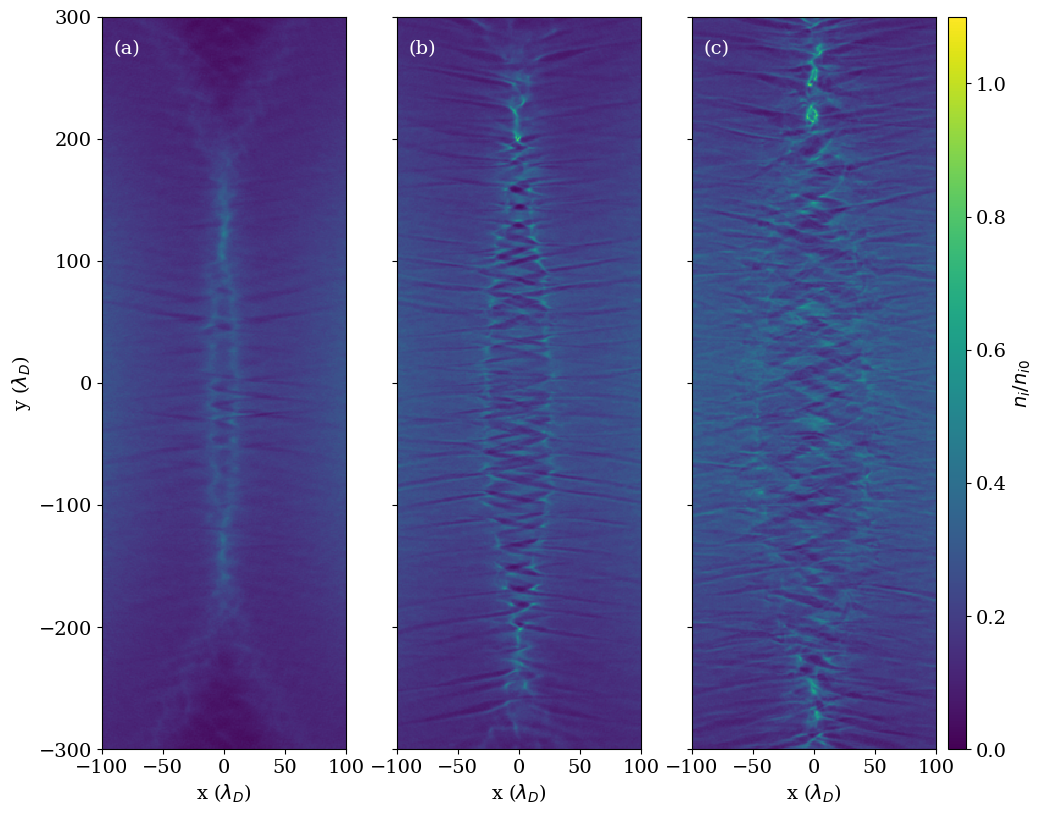}
    \caption{Ion density $n_i/n_{i0}$ of Simulation 3 at times (a) $7000\omega_{pe}^{-1}$, (b) $8600 \omega_{pe}^{-1}$, and (c) $10200\omega_{pe}^{-1}$. All panels share the same color scale.}
    \label{Dens_sim3}
\end{figure*}

Figure~\ref{Dens_sim3} shows the ion density.
Figure~\ref{Dens_sim3}(a) corresponds to the time $7000\omega_{pe}^{-1}$. It shows an elliptic structure with the largest width $\approx 40\lambda_D$ near $y=0$. This structure changes into two converging circular hybrid structures for $|y| > 220\lambda_D$. 
The elliptic structure is what we would expect from hybrid structures that pass through each other. A larger curvature outside $|y| = 220\lambda_D$ than inside suggests that the hybrid structures have been slowed by their interaction with the piston formed by the ambient ions. 

Figure~\ref{Dens_sim3}(b) shows the ion density at time $8600\omega_{pe}^{-1}$ corresponding to Fig.~\ref{PS_sim3}(c). The elliptic structure expanded along $x$ and its left and right boundaries near $y=0$ coincide with the positions along $x$ of the velocity modulations of ion beam~1 and~2 in the phase space density distribution. Oblique density cavities, which resemble those in Fig.~\ref{Dens_sim1}(b), point at ion-ion instabilities between both ion beams and ambient ions. 

Figure~\ref{Dens_sim3}(c) shows the ion density at time $10200\omega_{pe}^{-1}$. The elliptic structure has widened as expected of the outer boundaries that convect with the rarefaction wave ions.

\section{\label{discussion}Discussion}

We addressed with PIC simulations the collision of circular rarefaction waves and hybrid structures in a collisionless plasma. A rarefaction wave was launched at a circular dense plasma cloud. It expanded in vacuum or in ambient plasma with a low density. By letting the rarefaction wave cross the periodic boundary and reenter on the other side of the simulation box, we could model the collision of two rarefaction waves launched by the same dense plasma cloud.  

Simulation~1 considered the expansion in vacuum. The exponential density decrease of the rarefaction wave and the linear increase of its mean velocity with distance from the rarefaction wave front matched theoretical predictions. 

Both rarefaction waves collided at the symmetry axis, which was located halfway along the separation vector and perpendicular to it. The separation vector connected the center of the dense plasma with itself across the periodic boundary. Both rarefaction waves interpenetrated and their ions formed an overlap layer, which broadened with time. The relative speed of both ion beams projected on the separation vector was above $3c_s$ at time $7000\omega_{pe}^{-1}$. No shocks formed in simulation~1 and the ions of both rarefaction waves interacted via an ion-ion instability.      

In Simulation~2, an ambient plasma with a density 50 times less than that of the dense plasma was added. The expanding rarefaction waves swept up the ambient ions. Hybrid structures, which are composed of an electrostatic shock for the ambient ions and a double layer for the rarefaction wave ions, formed. The hybrid structures transported the accelerated ambient ions to the symmetry axis, creating an ion slab with hot and dense ambient ions. This ion slab was oriented perpendicularly to the separation vector and stationary in the rest frame of the simulation box. Its thermoelectric field formed a fast moving perturbation in the rest frame of each rarefaction wave. It triggered the growth of an ion acoustic wave, which steepened and changed into an electrostatic shock. 
An electrostatic shock formed in each rarefaction wave and propagated in the direction of the rarefaction wave front. These two reverse shocks enclosed a shared downstream plasma. The shocks could not slow the inflowing rarefaction wave ions to a standstill in the box frame. The drift of the ion beams downstream of both shocks led to the development of an ion phase-space hole. 

Both reverse shocks propagated fast in the rest frame of their rarefaction wave, but their speed in the simulation box frame was low. Self-similarity implies that the mean velocity of the rarefaction wave at a fixed position in space decreases with time. Eventually, the inflow speed will become subsonic, leading to the collapse of the shock or its propagation away from the ion slab. In our simulation, the reverse shocks collapsed. 

A density reduction of the ambient plasma in Simulation 3 implied that not enough ambient material could be piled up to shock the rarefaction wave plasma. Although the velocity distribution of the rarefaction wave ions was modified, which led to ion density oscillations, the plasma could not be compressed to the value needed to form a shock. Once the ambient ions dispersed, the ion density oscillations convected with the rarefaction wave. This created an elliptical structure in the ion density distribution that expanded with time.

Our results bear potential relevance for laser plasma experiments. Ablating a solid target with an ultra-intense laser pulse creates a plasma cloud that is orders of magnitude smaller and denser than the dense cloud we used in our simulation. As the laser-generated plasma plume expands, its density decreases due to the loss of material to the rarefaction wave. Its hemispherical expansion further implies a faster decrease with distance than in our 2D simulation. The rarefaction wave density decreases rapidly and becomes comparable to that of the ionized residual gas at a large enough distance from its source. When the expansion of the plasma plume is not too fast, on the order of a few hundred km/s, the residual gas plays a major role. For similar plasma plumes, it is the residual gas that controls the formation of reverse shocks.

Ablating two targets simultaneously and selecting suitable values for their separation and the density of the residual gas may allow us to recreate two reverse shocks, which enclose a downstream plasma in the plane halfway between the two targets. This also suggests that symmetric experimental configurations similar to that of Ref. [3] may generate more complex structures than a simple pair of collisionless shocks. The thermoelectric field of the reverse shocks should be detectable, for example, for the proton radiography method\cite{borghesi_propagation_2002}.

\begin{acknowledgments}

The simulations were performed on TGCC on resources provided by the Grand Equipement National de Calcul Intensif (GENCI) through the project No. 12943. M. F. acknowledge financial support by the Grand Programme de Recherche LIGHT for her PhD and by the LIGHTS\&T Graduate Program. The computations were enabled by resources provided by the National Academic Infrastructure for Supercomputing in Sweden (NAISS), partially funded by the Swedish Research Council through grant agreement no. 2022-06725.

\end{acknowledgments}

\section*{Author declaration}

\subsection*{Conflict of interests}
The authors have no conflicts to disclose.

\subsection*{Author contributions}

\textbf{Margaux François}:
Conceptualization (equal)
Data curation (lead)
Formal analysis (lead)
Investigation (lead)
Methodology (equal)
Resources (lead)
Visualization (lead)
Writing - original draft (equal)

\textbf{Mark Eric Dieckmann}:
Conceptualization (equal)
Formal analysis (supporting)
Investigation (supporting)
Methodology (equal)
Supervision (equal)
Writing - review \& editing (equal)

\textbf{Lorenzo Romagni}:
Conceptualization (equal)
Investigation (supporting)
Writing -review \& editing (equal)

\textbf{Xavier Ribeyre}:
Investigation (supporting)
Supervision (equal)
Writing - review \& editing (equal)

\textbf{Emmanuel d'Humières}:
Investigation (supporting)
Supervision (equal)
Writing - review \& editing (equal)

\section*{Data Availability Statement}

The data that support the findings of this study are available from the corresponding author upon reasonable request.

%\nocite{*}
\bibliography{aipsamp}% Produces the bibliography via BibTeX.

\end{document}